\documentclass[aps,prl,twocolumn,groupedaddress,amsmath,amssymb]{revtex4}

\usepackage{amssymb}
\usepackage{amsmath}
\usepackage{graphicx}
\usepackage{subfigure}
\usepackage{textcomp}
\usepackage{color}
\usepackage{amsfonts}
\usepackage{bbold}
\usepackage{dsfont}
\usepackage{epsfig}
\usepackage{hyperref}

\begin{document}
\title{Gate controlled spin pumping at a quantum spin Hall edge}

\author{Awadhesh Narayan} 
\thanks{These authors contributed equally to this work.}
\author{Aaron Hurley\footnotemark[1]}
\author{Stefano Sanvito}
\affiliation{School of Physics and CRANN, Trinity College, Dublin 2, Ireland}

\date{\today}

\begin{abstract}
We propose a four-terminal device designed to manipulate by all electrical means the spin of a magnetic adatom positioned at the 
edge of a quantum spin Hall insulator. We show that an electrical gate, able to tune the interface resistance between a quantum spin 
Hall insulator and the source and drain electrodes, can switch the device between two regimes: one where the system exhibits 
spin pumping and the other where the adatom remains in its ground state. This demonstrates an all-electrical route to 
control single spins by exploiting helical edge states of topological materials.
\end{abstract}

\maketitle


In recent years topological insulators and their two-dimensional analogs, quantum spin Hall (QSH) insulators, have been 
intensively investigated~\cite{review-kane,review-zhang}. Much of the excitement is due to their time-reversal symmetry 
protected edge states, which are immune to non-magnetic disorder. At a given edge spins of opposite direction counterpropagate, 
giving rise to spin-momentum locked helical states. These carry dissipationless current and appear promising for future spintronic 
applications~\cite{spin-pesin}. Already a number of proposals for devices utilizing these spin-momentum locked states have 
appeared. These range from topological spin transistors~\cite{qi-spinaharonov} to magnetoresistive devices~\cite{burkov-magnetores}, 
just to name a few. In parallel with these developments, rapid strides have been made in the precise control and manipulation of 
single atomic spins on insulating substrates by means of a Scanning Tunneling Microscope (STM) tip~\cite{Heinrich2004,Loth1}. 
Indeed it is now possible to measure the conductance spectra of these single atomic spins on nonmagnetic substrates at a rather
low temperature and in high magnetic fields. Advances in low-temperature STM have also allowed the detection of inelastic spin 
excitations in such setups and this has been termed spin-flip inelastic electron tunneling spectroscopy (sf-IETS).     
      

Inspired by the rapid developments in the two fields, we have showed in a recent study that the spin of a magnetic adatom 
could be manipulated by using the helical edge states of a QSH insulator~\cite{sanvito-IETS}. We found that, depending on the 
current intensity, electron transport presents two regimes: i) for a small current density there is suppression of the inelastic component 
of the conductance spectrum and, in contrast, ii) when the current density is large the conductance steps characteristic of the inelastic 
transitions re-appear. In the second regime spin angular momentum is pumped into the magnetic adatom, so that single spin 
manipulation by electrical means, a spintronics holy grail, is achieved. Here we extend the concept and propose a device where
the transition between the two regimes is also achieved electrically, by gating the region at the boundary between the QSH 
insulator and the source/drain electrodes. As such we will show that the QSH state can be coupled with sf-IETS to probe and 
manipulate single magnetic atoms without the need of a magnetic field or a spin-polarized STM tip.
\begin{figure}[bt]
\begin{center}
  \includegraphics[scale=1.0]{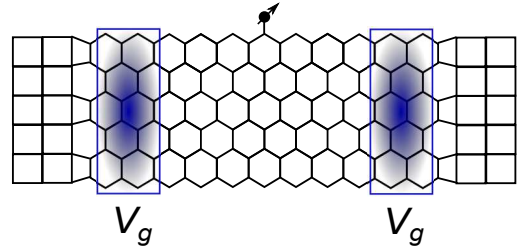}
  \caption{(Color online) A schematic diagram of the proposed device setup. A QSH insulator ribbon is connected to source and
  drain electrodes (represented by a simple square lattice) in a two terminal set up. A gate potential, $V_g$, is applied at the interface
  with both the source and the drain to control the interface resistance. The magnetic adatom with a local spin, $S$, is placed at one 
  of the edges of the ribbon. For the calculations we use a QSH insulator ribbon 7 sites long and 4 sites wide. The gate potential 
  is added to two layers of sites on either side of the ribbon. } \label{setup}
\end{center}
\end{figure}


The device setup is shown schematically in Fig.~\ref{setup}. It consists of two electrodes, described by a simple square lattice, 
sandwiching a QSH ribbon, which acts as scattering region. The magnetic adatom is positioned at one of the edges of the ribbon. 
Furthermore, we include two gate potentials at both the left-hand and right-hand side ends of the ribbon. The QSH ribbon is 
modeled by means of Kane-Mele Hamiltonian constructed on a honeycomb lattice~\cite{kane-mele},
\begin{align}\label{h-km} 
H_{\mathrm{KM}}=\varepsilon_0\sum_{i\alpha}\xi_i c^{\dagger}_{i\alpha}c_{i\alpha}&+t_1\sum_{\langle ij \rangle,\;\alpha}
c^{\dagger}_{i\alpha}c_{j\alpha}+\\ \nonumber
&+it_2\sum_{\langle\langle ij\rangle\rangle,\;\alpha\beta}\nu_{ij}c^{\dagger}_{i\alpha}[\sigma^z]_{\alpha\beta}c_{j\alpha}\:.
\end{align}
The first term is a staggered sublattice potential of strength $\varepsilon_0$ ($\xi_i=\pm 1$ for two sublattices of the honeycomb 
lattice). The second term is the nearest neighbor hopping with strength $t_1$ (we now set $t_1=1$). Finally the third term, which 
is crucial for the topological phase, is a complex next nearest neighbor hopping with strength $t_2$. This spin-orbit-mimicking term 
couples electrons' orbital motion to their spins via the $z$-component of the Pauli matrices ($\sigma^z$), with $\nu_{ij}=\pm 1$ 
having opposite sign for opposite directions of hopping. Although the spin-orbit term is too weak in graphene to observe this effect 
at realistic temperatures, recent proposals for silicene and its germanium analog have reported spin-orbit-driven bandgaps of 2.9 meV 
and 23.9 meV, respectively~\cite{silicene-yao}. Furthermore, Xu {\it et al.} have predicted two-dimensional Sn films to have spin-orbit 
gaps of 300 meV, which is comparable to that of the three-dimensional topological insulators currently known~\cite{sn-zhang}. These 
materials exhibit a low energy physics, which is well described by the Kane-Mele model.

In the presence of the gate voltages and of an adatom at site $i=I$, the total electronic Hamiltonian becomes
\begin{equation}
H_{\mathrm{el}}=H_{\mathrm{KM}}+V_g\sum_{i\epsilon\:\mathrm{gate},\:\alpha}c^{\dagger}_{i\alpha}c_{i\alpha}+\varepsilon_{I}\sum_{\alpha}c^{\dagger}_{I\alpha}c_{I\alpha}+t_I
\sum_{\langle Ii\rangle,\:\alpha}c^{\dagger}_{I\alpha}c_{i\alpha}\:,
\end{equation}
where in addition to $H_{\mathrm{KM}}$, we have the gate potential $V_g$, which is included via an additional on-site energy 
for the atoms in the gate region at the two ends of the ribbon. We have also included the on-site potential of the magnetic impurity, 
$\varepsilon_I$, and the hopping $t_I$ between the impurity site $I$ and its neighbor $i$ on the honeycomb lattice. For the magnetic 
impurity spin $\mathbf{S}$ we further introduce the Hamiltonian
\begin{equation}
H_{\mathrm{sp}}=DS^2_z\:;\;\;\;\;\;\;\;
H_{\mathrm{el-sp}}=J_{\mathrm{sd}}\sum_{\alpha\beta}c^{\dagger}_{I\alpha}[\boldsymbol{\sigma}]_{\alpha\beta}c_{I\beta}\cdot\mathbf{S}\:
\end{equation}
where $H_{\mathrm{sp}}$ describes a uniaxial anisotropy (along $z$, which is the direction normal to the plane of the QSH ribbon) 
with strength $D$. The coupling of the conduction electrons and the impurity spin is through the term $H_{\mathrm{el-sp}}$, with 
magnitude $J_{\mathrm{sd}}$. This Heisenberg-type coupling is the so-called $s$-$d$ model~\cite{sdmodel}.

The many-body problem introduced above is solved by perturbation theory in $H_{\mathrm{el-sp}}$. Here we outline only briefly the perturbation 
expansion needed to calculate the interacting self-energy and the spin state of the adatom, and we refer the readers to Refs.~\cite{Hurley1,Hurley3} 
for more mathematical details. The single-particle Green's function in the many-body ground state can be expanded to the $n$-th order in the 
interaction Hamiltonian. In this work we have truncated the expansion in the electron and spin propagator to the second-order. Then, we use 
Dyson's equation to calculate the interacting self-energy, which are then converted to real-time quantities by using Langreth's theorem. The 
dependence of the self-energy on the energy is then obtained by Fourier transform. The population of the spin states is evaluated from a 
master equation~\cite{sanvito-IETS}, which derives from the equation of motion for the spin propagator. Finally the net magnetization of the 
impurity spin is obtained by averaging over the population of all states.

In order to calculate the current we use the Landauer-B\"uttiker approach ~\cite{Buttiker} implemented in terms of the non-equilibrium 
Green's function (NEGF) method~\cite{Datta} and the conductance is then extracted as a numerical derivative. The retarded Green's 
function for the scattering region is given by, 
$G^r=[(E+i0^+)I-H_{\mathrm{el}}-\Sigma_\mathrm{L}-\Sigma_\mathrm{R}-\Sigma_{\mathrm{int}}]^{-1}$, where $\Sigma_\eta$ ($\eta=$~L, R) 
are the self-energies of the left-hand and right-hand side leads. The electron-spin coupling is taken into account via the interacting 
self-energy $\Sigma_{\mathrm{int}}$. Finally, the current can be calculated as
\begin{equation}
 I=\int_{-\infty}^{\infty}\frac{dE}{2\pi}[\Sigma^{<}(E)G^{>}(E)-\Sigma^{>}(E)G^{<}(E)],
\end{equation}
where $\Sigma^{\gtrless}$ ($G^{\gtrless}$) are the lesser and greater self-energies (Green's functions).


\begin{figure}[tb]
\begin{center}
  \includegraphics[scale=1.0]{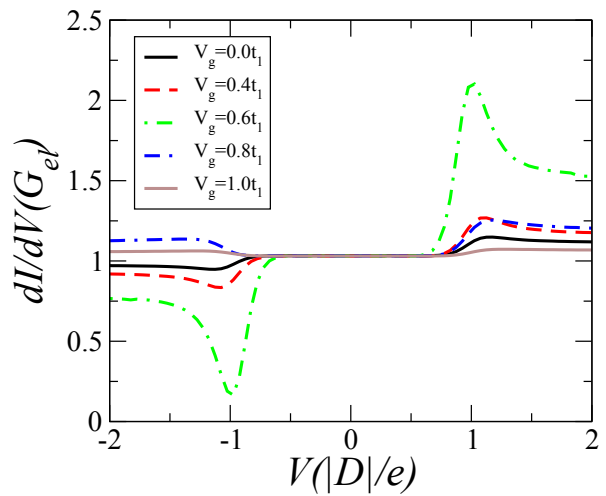}
  \caption{(Color online) Normalized conductance trace as a function of the source/drain voltage at different values of the applied gate 
  voltage for a $S=1$ impurity spin. Note that increasing the gate voltage beyond $V_g=0.6t_1$ allows us to crossover to a regime where 
  the current is reduced to a point at which the conductance steps are suppressed. The curves have been aligned vertically for ease 
  of comparison.} \label{vgate}
\end{center}
\end{figure}
We begin our analysis by looking at the conductance spectra (normalized to its $V=0$ value, $G_{el}$) at different values of the 
gate voltage, $V_g$. The results are presented in Fig.~\ref{vgate}. For the calculations we set $t_{\eta}=t_{\eta\:\mathrm{SR}}=t_{1}$ 
and $t_2=0.1t_1$ ($t_{\eta}$ is the hopping parameter in the electrodes and $t_{\eta\:\mathrm{SR}}$ is the coupling between the 
electrodes and the scattering region). These parameters keep the ribbon in the topological phase and the coupling between the 
leads and the scattering region is fixed to a large value. The choice of parameters, $\varepsilon_I=J_{\mathrm{sd}}=t_1/2$ and $t_I=t_1/4$, 
guarantees that perturbation theory can be used for the spin $S=1$. The spin degeneracy is lifted by introducing an axial 
anisotropy $D=-10^{-3}\:t_1$. This is equivalent to a temperature of around 12~K (assuming a realistic value of $t_{1}=1 eV$). 
In general, devices based on single atom anisotropies are low temperature devices expected to work around few tens of kelvin~\cite{Heinrich2004,Loth1}.

At $V_g=0$, i.e. when no gate voltage is applied, there is a conductance step seen at the energy corresponding to the first spin 
excitation of the system. This is the $|-1\rangle\rightarrow|0\rangle$ ($|-1\rangle\rightarrow|0\rangle$) spin transition 
for positive (negative) source/drain voltage, where $|S_z\rangle$ is the adatom spin third component~\cite{sanvito-IETS}. Such 
transition is detectable because the electrode-ribbon coupling strength is large and the current is intense. As we increase 
the gate voltage we observe an enhancement in the intensity of the conductance step. This continues utill $V_g$ reaches a 
value of $0.6t_1$. Beyond such critical gate voltage there is a drop in the conductance step at $V_g=0.8t_1$ and the inelastic 
conductance becomes suppressed. This suppression continues for higher values of the gate voltage. 

These observations can be readily understood in terms of the changing the interface resistance between the ribbon and the electrodes. 
At low values of $V_g$, there is a large current flowing from the leads to the ribbon and thus interacting with the magnetic adatom. 
For $V=0$ the spin on the adatom is in an equal superposition of $|+1\rangle$ and $|-1\rangle$ states. As the bias reaches $|D|/e$ 
the excitation to the $|0\rangle$ state is possible by spin-flip of the incoming electron. Since the current is large the impurity spin 
$S$ is not allowed to relax back to the ground state and thus the incident electrons can also induce transitions from $|0\rangle$ to 
$|+1\rangle$ and so be transmitted without getting backscattered. 

In the case of high interface resistance as engineered by increasing the gate voltage, there is a strong suppression of the 
conductance steps. In this scenario since the current density is small, the impurity spin can relax back to the ground state after 
the spin-flip event. This means that the incident right-going electrons, which are up spin polarized at the upper edge, will always 
encounter the impurity spin in either the $|+1\rangle$ or the $|-1\rangle$ state. A spin-flip event will reverse the electron's spin 
and since there are no down spin channels going right at the upper edge, the electron will be backscattered. Thus, the helicity 
of the QSH edge states leads to a suppression of the inelastic conductance steps at low currents~\cite{sanvito-Andreev}.
  
\begin{figure}[bt]
\begin{center}
  \includegraphics[scale=1.0]{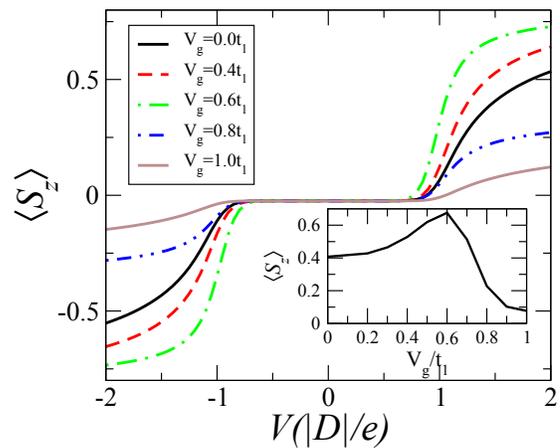}
  \caption{(Color online) Average magnetization, along the direction perpendicular to the ribbon plane, as a function of bias 
  voltage. Curves at different voltages are plotted showing gate control over the magnetization of the impurity spin. In the inset
  we report the magnetization as a function of gate voltage for a source/drain voltage of $V=1.5~|D|/e$.} \label{av-mag}
\end{center}
\end{figure}

Finally in Fig.~\ref{av-mag} we plot the average magnetization of the device at different gate voltages. If no gate voltage is applied 
then a change from zero to a finite magnetization occurs at $V=\pm|D|/e$, corresponding to the allowed spin excitation. Note that 
the direction of magnetization is opposite for opposite bias directions. As we now increase the gate voltage, the net magnetization 
increases, tending towards unity. This continues until $V_g=0.6t_1$ beyond which it drops rapidly. For higher values of $V_g$ the 
system remains closer to zero magnetization indicating the absence of spin pumping. The inset of Fig.~\ref{av-mag} traces the 
magnetization at $V=1.5~|D|/e$ as a function of the gate voltage. Note that the magnetization is always less than $\pm 1$, due to 
finite size of the QSH ribbon. 

In conclusion, we have studied the spin pumping of a quantum spin at a topological insulator edge. In particular we have
proposed a device concept in which an electrostatic gate in a two-terminal device allows the device switching between 
two regime. In the first inelastic conductance steps are suppressed, while in the second they are allowed. This second regime 
offers the possibility of manipulating impurity spins using the helical edge states of a QSH insulator. Thus, we have proposed 
a complete all-electrical device which allows control of a quantum spin without using any magnetic field.  


This work is supported by Irish Research Council (AN) and Science Foundation of Ireland (AH, grant No. 07/IN.1/I945). 
Computational resources have been provided by Trinity Center for High Performance Computing. We acknowledge 
illuminating discussions with Cyrus Hirjibehedin, Ivan Rungger and Nadjib Baadji.

\end{document}